# TIME EVOLUTION OF NON-LETHAL INFECTIOUS DISEASES: A SEMI-CONTINUOUS APPROACH.

A. Noviello

*Dipartimento di Matematica ed Applicazioni "Renato Caccioppoli"*

*Università degli Studi di Napoli "Federico II"*

*I-80100 Napoli, Italy*

F. Romeo

*Dipartimento di Fisica "E. R. Caianiello"*

*Università degli Studi di Salerno*

*I-84081 Baronissi (SA), Italy*

R. De Luca

*INFM and DIIMA, Università degli Studi di Salerno*

*I-84084 Fisciano (SA), Italy*

## ABSTRACT

A model describing the dynamics related to the spreading of non-lethal infectious diseases in a fixed-size population is proposed. The model consists of a non-linear delay-differential equation describing the time evolution of the increment in the number of infectious individuals and depends upon a limited number of parameters. Predictions are in good qualitative agreement with data on influenza.

## I INTRODUCTION

The dynamical laws governing the spreading of infectious diseases among a closed population are of interest mainly in the field of medical science. However, owing to the stirring of public opinion generated by the diffusion of lethal viruses, social issues are also to be considered. Indeed, much attention has been lately devoted to the study of the HIV virus, given its strongly lethal character, so that traditional mathematical models [1, 2] have been modified according to the characteristic features of this illness [3, 4]. More recently, SARS has had a great social and economic impact in the whole world [5].

Even the most common non-lethal viral infections as, for example, influenza, are cause of some social distress, given their periodic appearance and their potentiality of being harmful to the weak exposed population. Classic models [6] give a fairly good description of the time evolution of infectious diseases. Most of the assumptions made in this traditional type of approach greatly simplify the analysis of the complex problem of the spreading of such illnesses among a certain number of individuals. However, in order to give account of the great variety of responses to the same viral infection and to distinguish among different social habits of different individuals, a network approach can be adopted [7, 9]. The topological issues addressed by the network approach are very important *per se*, since they find application in other fields, such as, for example, sociology [10]. In addition, they are useful in describing, more realistically, the problem and in defining the limits of validity of traditional models.

In the present work, we propose, under the same basic hypotheses of traditional models, a modified SIR (Susceptible-Infectious-Recovered) model. The time evolution of non-lethal infectious illnesses in a community of highly mobile individuals or, equivalently, in a population with different species uniformly distributed over the observation landscape, will be analyzed.

The paper will be organized in two main parts. In the first part we reconsider the equations governing the time evolution of SIR models by assuming that an individual, infected at time $t$, recovers after an interval of time $\tau$, which is taken to be the same for every population member. The interaction between the S-I species is taken to be regulated by a constant statistical parameter $\pi$, while other horizontal cross-interactions are neglected. The time interval in which the S- species population is monitored, is assumed to be small, in such a way that the total number of individuals can be thought to be constant over the entire observation period.

In the second part of the present paper the resulting non-linear delay-differential equations [11] for the increment in the number of infected individuals are solved by standard numerical routines. The duration of the disease as a function of a rescaled interaction parameter $\hat{\pi}$ is plotted for different initial numbers of infectious individuals $\tilde{\mu}_0$ and for different population sizes. Endemic-like and epidemic regimes for the infectious disease, as also found by means of traditional approaches [6], are defined by introducing the cross-over value $\hat{\pi}_c$ of the effective interaction parameter. Conclusions are drawn in the last section, where further developments of the modified model are also discussed.

## II THE MODEL

Let us consider a fixed size population of individuals, subdivided in three distinct species: Susceptible (S-species); Infectious (I-species); Recovered (R-species). Assume that the mobility of each individual is such to cause a complete interaction among all members of the community within one monitoring time interval $\Delta t$. Equivalently, one might also assume that the members of each species are uniformly distributed, at any time $t$, over the observation landscape. Under this hypothesis we can take the number of new infections $\Delta\mu_A$, due to a non-lethal virus, to be given by the following relation [1]:

$$\Delta\mu_A = \pi\mu(t)\sigma(t)\Delta t, \qquad (1)$$

where $\mu(t)$ and $\sigma(t)$ are the number of infectious and susceptible individuals at time $t$, respectively, and where $\pi$ is the effective infection rate. A slight different meaning is to be attributed to the statistical exchange parameter $\pi$, depending whether we are observing highly mobile individuals or uniformly distribute species over the entire landscape. In the present model, indeed, we shall introduce a recovery time interval (or infectious period) $\tau$, taken to be equal for all individuals. This quantity acts as a reference time scale, so that, in the case of highly mobile individuals, $\pi$ is representative of the effectiveness of the interaction between the S- and the I- species in a time $\tau$, while it accounts also for the number of these interactions in the same interval of time in the case of uniform distribution of the three species. By now defining $\rho(t)$ as the total number of recovered individuals at time *t*, we may write:

$$N = \mu(t) + \sigma(t) + \rho(t), \tag{2}$$

where *N* is the total constant number of members in the community.

Let us now subdivide the time interval $[0,+\infty)$ into contiguous sub-intervals $I_n = [(n-1)\tau, n\tau]$, where *n* is a positive integer. For $t \in I_n$, we define the number of individuals belonging to the S-, I- and R- species, respectively, as follows:

$$\sigma_n(t) = \sigma_{n-1}((n-1)\tau) - a_n(t), \tag{3a}$$

$$\mu_n(t) = \mu_{n-1}((n-1)\tau) + a_n(t) - a_{n-1}(t-\tau), \tag{3b}$$

$$\rho_n(t) = \rho_{n-1}((n-1)\tau) + a_{n-1}(t-\tau), \tag{3c}$$

where the function $a_n(t)$ counts the number of infected individuals in $I_n$, up until time *t*, starting with a null initial count at $t = (n-1)\tau$, i.e., for each lower bound of the interval $I_n$, so that:

$$a_n((n-1)\tau) = 0, \tag{4a}$$

$$a_n(n\tau) = \mu_n(n\tau). \tag{4b}$$

Notice that the term $a_{n-1}(t-\tau)$ counts the number of individuals that get infected in the interval $[(n-2)\tau, (t-\tau)] \subseteq I_{n-1}$. These individuals recover in the time interval $[(n-1)\tau, t] \subseteq I_n$. In order to describe the dynamics of the population of the I-species, we also consider the number of individuals which recover in the interval $\Delta t$, so that the variation in the number of infected $\Delta\mu_n$ in this time interval is given by the following expression:

$$\Delta\mu_n = \pi\mu_n(t)\sigma_n(t)\Delta t - [a_{n-1}(t-\tau+\Delta t) - a_{n-1}(t-\tau)], \tag{5}$$

where the second term represents the number of individuals getting infected in the interval $[t-\tau, t-\tau+\Delta t] \subseteq I_{n-1}$ and recovering in the interval $[t, t+\Delta t] \subseteq I_{n-1}$.

For $\Delta t << \tau$, we thus have:

$$\dot{\mu}_n(t) = \pi\mu_n(t)\sigma_n(t) - \dot{a}_{n-1}(t-\tau), \tag{6}$$

where the dot on top of the variable stands for its time derivative and where $n \geq 2$, $n$ being an integer. For $n = 1$, Eq. (6) reduces to the following $\dot{\mu}_1(t) = \pi\mu_1(t)\sigma_1(t)$, being $a_n(t)$ defined only for positive integer indices. Eq. (6) thus shows that the dynamical problem can be described by means of a collection of non-linear coupled differential equations, each one defined on a time interval of length $\tau$. However, we can postulate the existence of a function $m : \Re_0^+ \to \Re_0^+$, which is continuous over the time domain $\Re_0^+$ and whose restriction on the time intervals $I_n$ is given by the following expression:

$$m_n(t) = m_{n-1}((n-1)\tau) + a_n(t). \tag{7}$$

Notice that the requirement of continuity of the global function $m(t)$ is fulfilled if the functions $m_n(t)$ are continuous over $I_n$. Indeed, by Eq. (4), we have that, at the common point $t_n = (n-1)\tau$ of the contiguous time intervals $I_{n-1}$ and $I_n$, Eq. (7) gives $m_n((n-1)\tau) = m_{n-1}((n-1)\tau)$. Continuity of the function $m(t)$ can thus be fully proven by exhibiting the single functions $m_n(t)$, after we have solved the problem.

Before finding and solving the differential equation for the function $m(t)$, let us consider what follows. First of all notice that, since $\mu_n(n\tau) = a_n(n\tau)$ by Eq. (4b), for $n \geq 2$ we can write:

$$m_n(n\tau) = m_{n-1}((n-1)\tau) + a_n(n\tau) = m_{n-1}((n-1)\tau) + \mu_n(n\tau). \tag{8}$$

We might extend the validity of the above expression to $n = 1$, by defining the following zero-index functions in the interval $I_0 = [-\tau, 0]$:

$$m_0(t) = \mu_0(t) = \begin{cases} \tilde{\mu}_0 & \text{for } t = 0 \\ 0 & \text{for } -\tau \le t < 0 \end{cases}, \tag{9}$$

$$\sigma_0(t) = \begin{cases} N - \tilde{\mu}_0 & \text{for } t = 0 \\ 0 & \text{for } -\tau \le t < 0 \end{cases}, \tag{10}$$

$$\rho_0(t) = 0 \qquad \text{for } -\tau \le t \le 0 . \tag{11}$$

where $\tilde{\mu}_0$ is the initial number of infected individuals. At $t = 0$ these functions reproduce the correct initial conditions of the problem and are compatible with the form of the differential equation given by Eq. (6) for $n = 1$.

We now find the time evolution of the functions $m_n(t)$ for $n \ge 1$. Making use of Eqs. (3a-c), and considering the definition of the zero-index functions above, we first rewrite Eq. (6) in terms of the functions $a_n(t)$ for $n \ge 1$ as follows:

$$\dot{a}_n(t) = \pi[\mu_{n-1}((n-1)\tau) + a_n(t) - a_{n-1}(t-\tau)][\sigma_{n-1}((n-1)\tau) - a_n(t)]. \tag{12}$$

By now solving for $a_n(t)$ in Eq. (7) in terms of the functions $m_n(t)$ and by substituting in Eq. (12), we have:

$$\dot{m}_n(t) = \pi[m_n(t) - m_{n-1}(t-\tau)][N - m_n(t)]. \tag{13}$$

In deriving the above equation we have made use of the following relations:

$$m_{n-1}((n-1)\tau) - m_{n-2}((n-2)\tau) - \mu_{n-1}((n-1)\tau) = 0 , \tag{14}$$

$$\sigma_{n-1}((n-1)\tau) + m_{n-1}((n-1)\tau) = \sigma_n(n\tau) + m_n(n\tau) = N . \tag{15}$$

Eq. (14) follows directly from Eq. (8), while Eq. (15) can be proven to be valid since, by definition of $\sigma_n(t)$ (Eq. (3a)) and of $m_n(t)$ (Eq. (7)) , we may write:

$$\sigma_n(n\tau) + m_n(n\tau) = \sigma_{n-1}((n-1)\tau) + m_{n-1}((n-1)\tau) = \sigma_n(t) + m_n(t). \tag{16}$$

Given that the above relation should be valid also for $n = 1$ and for $t = 0$, we have

$$\sigma_n(t) + m_n(t) = \sigma_0(0) + m_0(0) = N , \tag{17}$$

thus proving the statement in Eq. (15).

We can now built up a global solution $m(t)$, by solving the dynamical equation given in Eq. (13) in all contiguous intervals $I_n = [(n-1)\tau, n\tau]$, for $n = 1, 2, 3, \ldots$, starting from $n = 1$ and proceeding for increasing integer values of *n*. For each step, the functions $\sigma_n(t)$, $\mu_n(t)$, and $\rho_n(t)$ which can be considered as the restrictions of global functions $\sigma(t)$, $\mu(t)$, and $\rho(t)$ on the interval $I_n$, can be found by the following inverse relations:

$$\sigma_n(t) = N - m_n(t), \tag{18a}$$

$$\mu_n(t) = m_n(t) - m_{n-1}(t-\tau), \tag{18b}$$

$$\rho_n(t) = m_{n-1}(t-\tau). \tag{18c}$$

## III SOLUTIONS AND RESULTS

In the present section we shall proceed to solve the non-linear delay-differential equations for the functions $m_n(t)$ (Eq. (13)) in sequence and shall discuss the results. Let us then start by the first time interval $I_1$. In this case we need to solve the logistic equation

$$\dot{m}_1(t) = \pi m_1(t)[N - m_1(t)], \tag{19}$$

with initial conditions $m_1(0) = \tilde{\mu}_0$. We can thus immediately write down the solution to Eq. (19) [12]

$$m_1(t) = \frac{\tilde{\mu}_0 N}{\tilde{\mu}_0 + (N - \tilde{\mu}_0)e^{-\pi N t}}, \tag{20}$$

while the functions $\sigma_1(t)$, $\mu_1(t)$, and $\rho_1(t)$ can be found by means of Eqs. (18a-c).

In the time interval $I_2$, the differential equation in Eq. (13) takes the following form

$$\dot{m}_2(t) = \pi[m_2(t) - m_1(t-\tau)][N - m_2(t)]. \tag{21}$$

Solution to the above equation can now be sought by utilizing the expression for $m_1(t)$ in Eq. (20) and by using the initial condition

$$m_2(\tau) = m_1(\tau) = \frac{\tilde{\mu}_0 N}{\tilde{\mu}_0 + (N - \tilde{\mu}_0)e^{-\pi N \tau}}. \tag{22}$$

The same procedure can be used, iteratively, for $n = 3, 4, 5, \ldots$, until a satisfactory picture of the global solution $m(t)$ is obtained.

Before presenting the outcome of the numerical integration, we might notice that, by defining the following normalized quantities:

$$\xi = \frac{t}{\tau}; \quad \widehat{m} = \frac{m}{N}; \quad \widehat{\pi} = \pi N \tau , \tag{23}$$

Eq. (13) can be cast in the form

$$\frac{d}{d\xi}\widehat{m}_n(\xi) = \widehat{\pi}\left[\widehat{m}_n(\xi) - \widehat{m}_{n-1}(\xi - 1)\right]\left[1 - \widehat{m}_n(\xi)\right]. \tag{24}$$

In Eq. (24) the role of the parameters $\pi$, N, and $\tau$ in the present model is clarified. Indeed, the dynamical behaviour of the system appears to be independent from the infectious period $\tau$ and from the population size N, since, from Eq. (24), the quantities $\tau$ and N simply rescale the infection rate $\pi$. We thus expect a universal behaviour of the observed experimental data, independently from the population size and for the recovery time interval $\tau$. The only significant parameters in the universal curves are the effective interaction parameter $\widehat{\pi}$, which in traditional models is referred to as contact number, and the initial percentage of infected individuals $p_0 = \frac{\widetilde{\mu}_0}{N}$. As far as the duration of the infection is concerned, however, we shall see, in what follows, that the population size does play a role in determining the lower limit of the function $\widehat{\mu}_n(t) = \frac{\mu_n(t)}{N}$. We also notice that, by definition of the number of individuals in each species in Eqs. (18a-c) and by Eq. (24), we get:

$$\widehat{\sigma}(\xi) = \widehat{\sigma}(0) e^{-\widehat{\pi}\int_0^{\xi}\widehat{\mu}(\lambda)d\lambda}, \tag{25}$$

which gives the residual percentage of susceptible individuals $\widehat{\sigma}(\xi)$ not yet infected at normalized time $\xi$ in terms of the total count of infected individuals, normalized to the $N$, up to $\xi$.

The integration of Eq. (24) is performed by standard numerical routines. In Figs. 1a-b we report, for two different values of $\widehat{\pi}$, the time evolution of the function $\widehat{m}(\xi)$, in order to detect its continuity properties. In Figs. 2a-b and in Fig. 3 we show the functions $\mu$, $\sigma$, and $\rho$ normalized to *N*, which we indicate as $\hat{\mu}$, $\hat{\sigma}$, and $\hat{\rho}$, respectively, versus the normalized time $\xi = \frac{t}{\tau}$, for $p_0 = 0.01$. In Figs. 2a-b we have chosen a statistical parameter $\hat{\pi}$ equal to *0.95*, while in Fig. 3 we have taken $\hat{\pi} = 3.5$. From these time-evolution curve, we first notice, starting with Figs. 1a-b, that $\hat{m}(\xi)$ is a monotonically increasing function and attains characteristic s-shape for $\hat{\pi} > 1$. In fact, in Fig. 1a, obtained for $p_0 = 0.01$ and $\hat{\pi} = 0.95$, no clear inflection point can be detected in the curve, while in Fig. 1b, obtained for $p_0 = 0.01$ and $\hat{\pi} = 3.5$, an inflection point can be clearly seen in the interval $1 < \xi < 2$. We also notice that, in correspondence to this feature, the model predicts an endemic-like regime for $\hat{\pi} < 1$. In the particular case of Figs. 2a-b, where $\hat{\pi} = 0.95$, the infection is present in the community for a rather long time, while the number of infected individuals is low at any instant of time. A different behavior can be seen when $\hat{\pi} = 3.5$, for example, as in Fig. 3, where an epidemic regime is detected. Indeed, in this case, the percentage of infected individuals reaches rather high values, even though not all individuals get infected in the interval of duration of the disease.

An interesting result is found when we let the population size *N* vary. As said in the previous section, one can argue that the dynamics of the phenomenon is not affected by the parameter *N*, since this extensive quantity collapses into the rescaled parameter $\hat{\pi}$ after normalization. One cannot exclude, however, that observable quantities, like, for example, the duration of the malady expressed in terms of normalized time, would be independent of *N*. Indeed, by considering $\hat{\pi}$ fixed, if *N* is increased, one needs to decrease the product $\pi\tau$, which describes the average number of infected people in the normalized time interval $\hat{I}_n = [(n-1), n]$, $n \geq 1$. In this way, the duration of the malady spreads over a longer normalized time interval. Numerically this result is obtained by adjusting the lower level of the normalized number of infected individuals when deciding at what instant of time the disease ends. Indeed, let us suppose that the population size is $N_0$, then the duration of the malady is determined by intersecting the curves $\hat{\mu} = \hat{\mu}(\xi)$ and $\hat{\mu} = \frac{1}{N_0}$, and by looking at the instant of time at which intersection of the two curves occurs. Therefore, if we consider the parameter $\hat{\pi}$ fixed, but consider two population sizes, say $N_1$ and $N_2$, with $N_1 < N_2$, then we need to intersect the curves $\hat{\mu} = \frac{1}{N_1}$ and $\hat{\mu} = \frac{1}{N_2}$ with the same solution $\hat{\mu} = \hat{\mu}(\xi)$ of the dynamical equation (24). Let us assume that intersections occur, respectively, at normalized times $\xi_1$ and $\xi_2$. Given that the curve $\hat{\mu} = \hat{\mu}(\xi)$ decreases to zero for $\xi >> 1$, one can finally argue that $\xi_1 < \xi_2$, as it appears from Figs. 4a-b, where the duration of the infection versus the value of the statistical parameter $\hat{\pi}$ is shown for $p_0 = 0.01$ and $p_0 = 5 \cdot 10^{-3}$, respectively. In these figures various values of the population size *N* have been used and we notice that, for increasing values of the population number, by keeping fixed $\hat{\pi}$ and $p_0$, the duration of the infection increases.

In Fig. 5a and Fig. 5b the abscissa and the ordinate, respectively, of the maximum points in the $\hat{\mu}$ vs. $\xi$ curves are reported as a function of $\hat{\pi}$ for three different values of the initial percentage of infections $p_0$. In Fig. 5a a clear landmark of the two different regimes, the endemic-like regime and the epidemic one, is the abrupt crossover from a plateau in the $\xi_{Max}$ vs. $\hat{\pi}$ curves to a monotonously decreasing behaviour. This crossover appears around the value $\hat{\pi} \cong 1$, coherently with what noted before. This type of behaviour is well-known from standard epidemiological theories [6], where the basic reproductive number $R_0$, defined as the average number of secondary infections caused by the introduction of a single infected individual into an entirely susceptible population, is considered. Epidemic outbreaks are expected to appear, according to traditional models, if $R_0 > 1$. As for the present model, the transition between the two regimes, clearly detected in the $\xi_{Max}$ vs. $\hat{\pi}$ curves shown in Fig. 5a, is also detectable in the $\hat{\mu}_{Max}$ vs. $\hat{\pi}$ curves reported in Fig. 5b. In the latter case, however, the rather sharp knee in the curves at low values of the initial percentage of infections disappears for higher values of $p_0$, meaning that there cannot be a clear definition of the two regimes if there is a relevant percentage of individuals infected at the time of invasion.

In order to see more closely how the cross-over value $\bar{\pi}_c$ varies with the initial percentage of infected individuals, in Fig. 6 we show a set of numerically evaluated points which give, for certain values of $p_0$, the value of the interaction parameter leading to transition from the endemic-like to the epidemic regime. The full-line curve is obtained by a best fit procedure and follows a square-root law.

As a final comment we notice that the bell-shaped curve for the percentage of infected individuals is in good qualitative agreement with observed data for influenza [13].

## IV CONCLUSION

We have developed an analytic model to describe the time evolution of non-lethal infectious diseases in a fixed-size population of *N* individuals. The population consists of S-, I-, R-species (respectively, susceptible, infected and recovered species). In the model a recovery time $\tau$ is introduced. This parameter acts as a delay time in the resulting non-linear delay-differential equations for the model. It has been noted that $\tau$ not only induces a natural way of normalizing the time variables, but also determines a time scan in the model, in such a way that a sequence of differential equations has to be found, one for each time interval $I_n = [(n-1)\tau, n\tau]$, with $n \geq 1$. In order to solve more adequately these equations, a continuous variable $m(t)$ is introduced. The time evolution of the restriction of $m(t)$ on the time interval $I_n = [(n-1)\tau, n\tau]$, denoted as $m_n(t)$, is described by the dynamical equation found for that interval of time. The number of individuals belonging to the S-, I-, and R-species are expressed in terms of the functions $m_n(t)$, for each time interval $I_n$, with $n \geq 1$. The complete evolution of the species is obtained by *gluing* these partial solutions together.

We notice that the present model is different from the classic SIR model [6] since, in Eq. (6) a time delay is explicitly considered in order to properly count the number of individuals which recover at time $t$ after having been infected at time $t-\tau$. This feature is useful to describe the spread of the illnesses at the time of invasion, in such a way that the only necessary parameters for the model are the effective contact rate $\hat{\pi}$ and the initial percentage of infected individuals $p_0$. This allows us to define a universal dynamics for these types of diseases in terms of a limited number of parameters, but does not exclude that the population size may play a role in determining observable quantities like, for example, the duration of the infection. Indeed, when the latter quantity is plotted against the parameter $\hat{\pi}$, one notices different quantitative behaviour in the curves for different values of *N*. This behaviour can be interpreted by recalling the definition of the parameter itself and appears because of the need of defining a lower level for the normalized value of the fraction of infected individuals, below which the infection ceases to be present in the community. A clear distinction between two types of regimes, epidemic and endemic-like, is attained by reporting the normalized value of the time at which the maximum in the curve of the percentage of infectious individuals appears. The cross-over value $\hat{\pi}_c$ of the effective interaction parameter is determined numerically in terms of the initial percentage of infected individuals $p_0$. The bell-shaped curves describing the time evolution of the number of infections in the epidemic regime are seen to be in good qualitative agreement with available data on influenza.

In future works we shall gradually increase the degree of complexity of this modified classic model, by first considering a SEIR model with a fourth species (Exposed or E-species), defining an incubation time $q$ for the particular illness considered. Successively, we shall take account of the lethal character of some infectious diseases by introducing an additional parameter giving the average percentage of individuals subsiding after infection and by allowing the population size to depend on time.

## REFERENCES

1. R. M. Anderson and R. M. May, Nature **280**, 361 (1979); R. M. Anderson and R. M. May, ibid. 280, 455 (1979).
2. R. M. Anderson and R. M. May, Nature **280**, 455 (1979).
3. J. M. Hyman, Jia Li, and E. A. Stanley, Math. Biosci. **155**, 77 (1999).
4. M. Y. Li, J. R. Graef, L. Wang, and J. Karsai, Math. Biosci. **160**, 191 (1999).
5. M M. Lipshitch, T. Cohen, B. Cooper, J. M. Robins, S. Ma, L. James, G. Gopalakrishna, S. K. Chew, C. C. Tan, M. H. Samore, D. Fisman, M. Murray, Science **300**, 1966 (2003).
6. H. W. Hethcote, SIAM Review **42**, 599 (2000).
7. M. E. J. Newman, Phys. Rev. E **66**, 016128 (2002).
8. R. Pastor-Satorras and A. Vespignani, Phys. Rev. E **63**, 066117 (2001).
9. R. M. May and A. L. Lloyd, Phys. Rev. E **64**, 066112 (2002).
10. D. J. Watts and S. H. Strogatz, Nature **393**, 440 (1998).
11. Thomas L. Saaty, *Modern Nonlinear Equations* (Dover Publications, New York, 1981).
12. W. E. Boyce and R. C. DiPrima, *Elementary Differential Equations and Boundary Value Problems* (John Wiley and Sons, New York, 1977).
13. Department of Health and Human Services, Centers for Disease Control and Prevention, *Weekly Report: Influenza Summary Update*, http://www.cdc.gov/flu/weekly/ (2004).

**FIGURE CAPTIONS**

1. Time dependence of the function $\hat{m}(\xi)$ for $p_0 = 0.01$ and for two different values of the effective interaction parameter : a) $\hat{\pi} = 0.95$; b) $\hat{\pi} = 3.5$.

2. a) Dynamical evolution of the normalized number of individuals belonging to the S-, I-, R-species for $\hat{\pi} = 0.95$ and for $p_0 = 0.01$. The variable on the vertical axis is representative of the functions $\hat{\mu}$, $\hat{\sigma}$ , and $\hat{\rho}$, which are respectively graphed with dashed, dash-dotted and full lines. b) a magnification of the lower portion of the graph in a).

3. Dynamical evolution of the normalized number of individuals belonging to the S-, I-, R-species for $\hat{\pi} = 3.5$ and for $p_0 = 0.01$. The variable on the vertical axis is representative of the functions $\hat{\mu}$, $\hat{\sigma}$ , and $\hat{\rho}$, which are respectively graphed with dashed, dash-dotted and full lines.

4. Duration *T* of the infection as a function of the value of the effective interaction parameter $\hat{\pi}$ for a) $p_0 = 0.01$ and for b) $p_0 = 5 \cdot 10^{-3}$. The curves are obtained for various values of the population size *N* (from bottom to top, alternating full and dot lines: $N = 10^3, 10^4, 10^5, 10^6$).

5. a) Abscissa and b) ordinate of the maximum points appearing in the time evolution curves of the percentage of infectious individuals. These values have been numerically determined and are reported in terms of the effective interaction parameter $\hat{\pi}$ for (bottom to top): a) $p_0 = 10^{-2}, 5 \cdot 10^{-3}, 10^{-3}, 5 \cdot 10^{-4}$; b) $p_0 = 10^{-4}, 5 \cdot 10^{-3}, 10^{-2}, 2.5 \cdot 10^{-2}, 5 \cdot 10^{-2}, 0.1, 0.2$.

6. Critical effective interaction parameter $\hat{\pi}_c$ as a function of the percentage of initially infected individuals. The points are numerically determined by reporting the values of $\hat{\pi}$ at which the $\xi_{\max}$ vs $p_0$ curves shows deviation from endemic-like behavior, characterized as a plateau in Fig. 5a. The full-line curve is determined through a best-fit procedure and is given by the following functional relation: $f(p_0) = 0.9934\left(1+\sqrt{p_0}\right)$.

**Fig. 1**

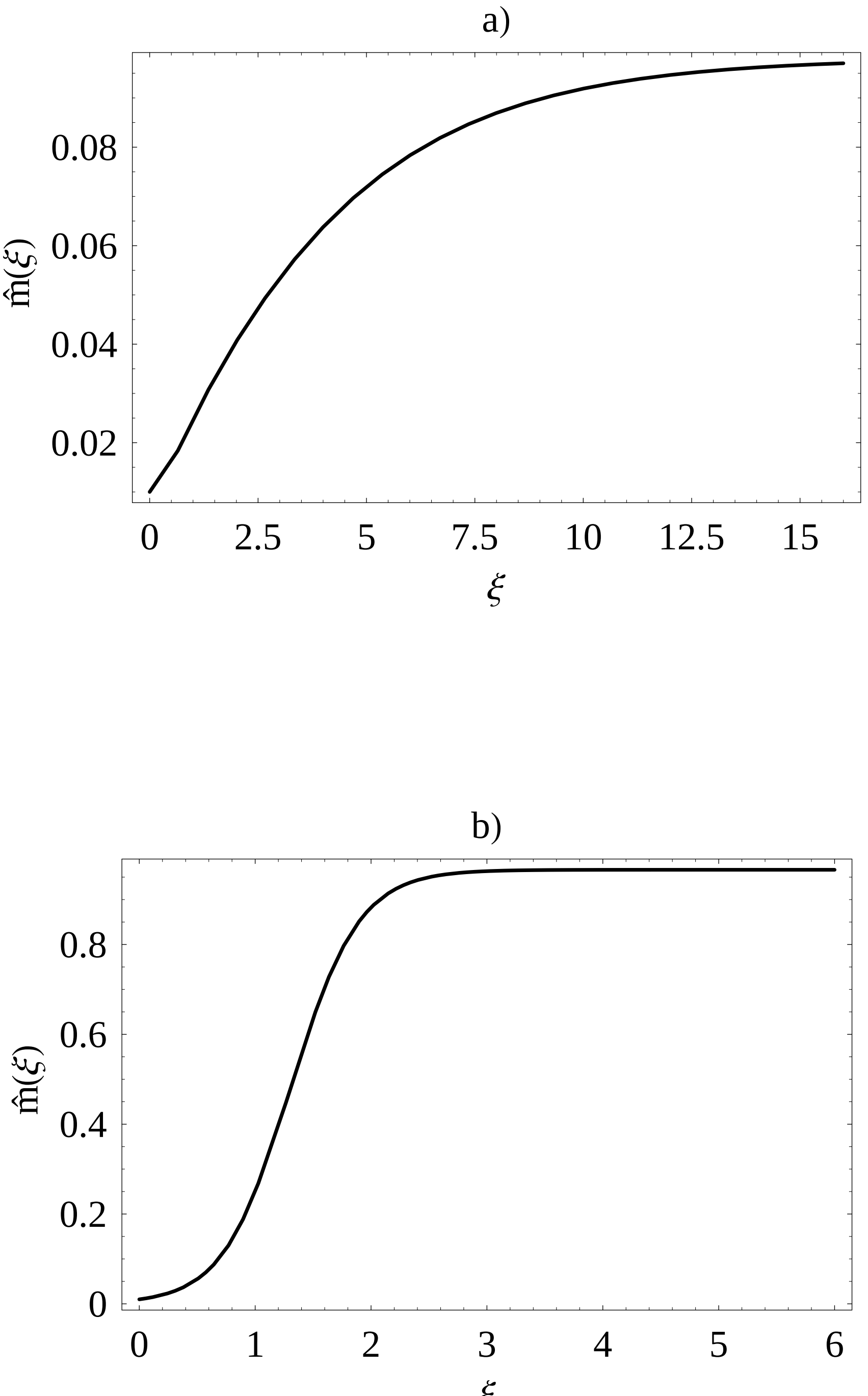

**Fig. 2**

a)

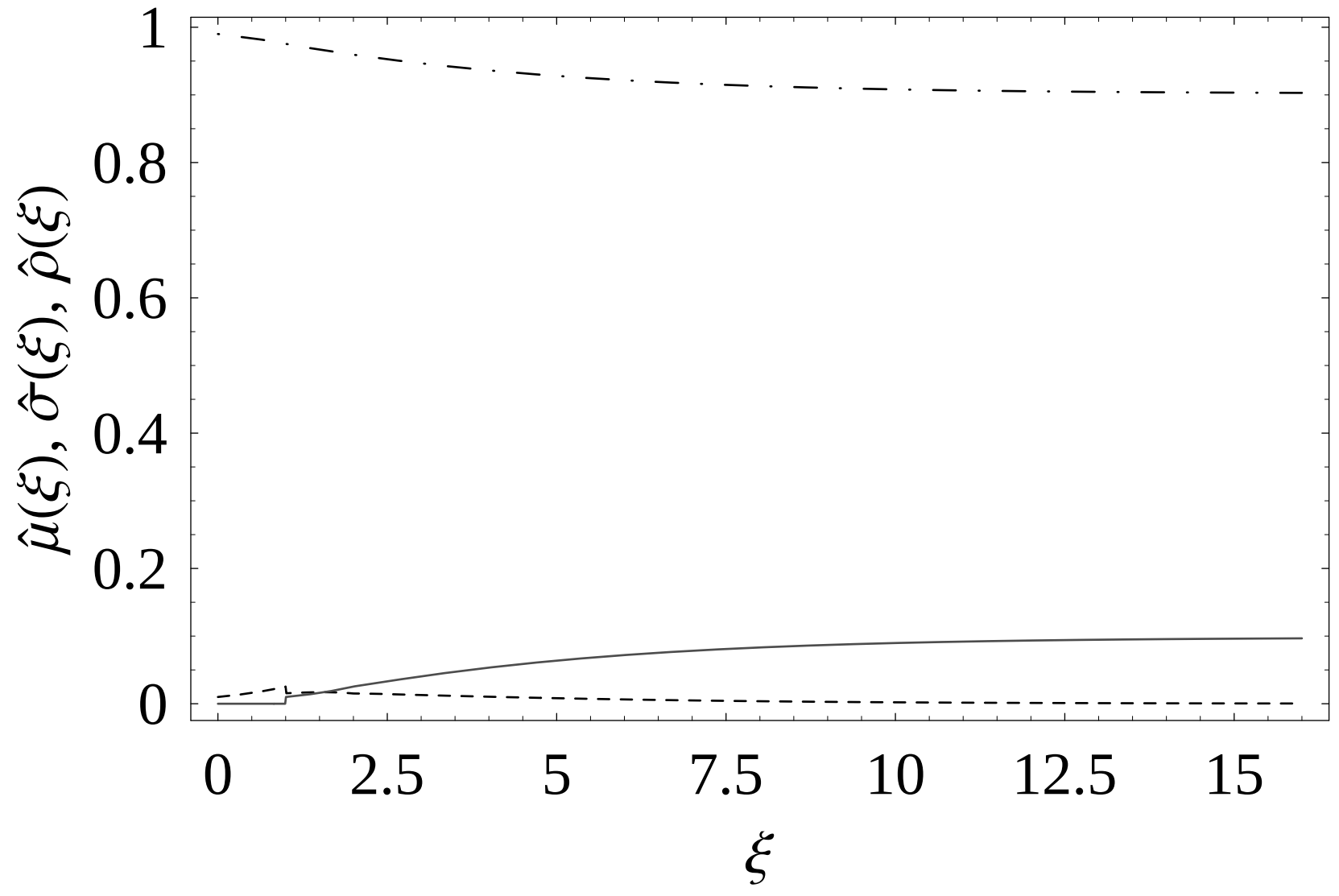

b)

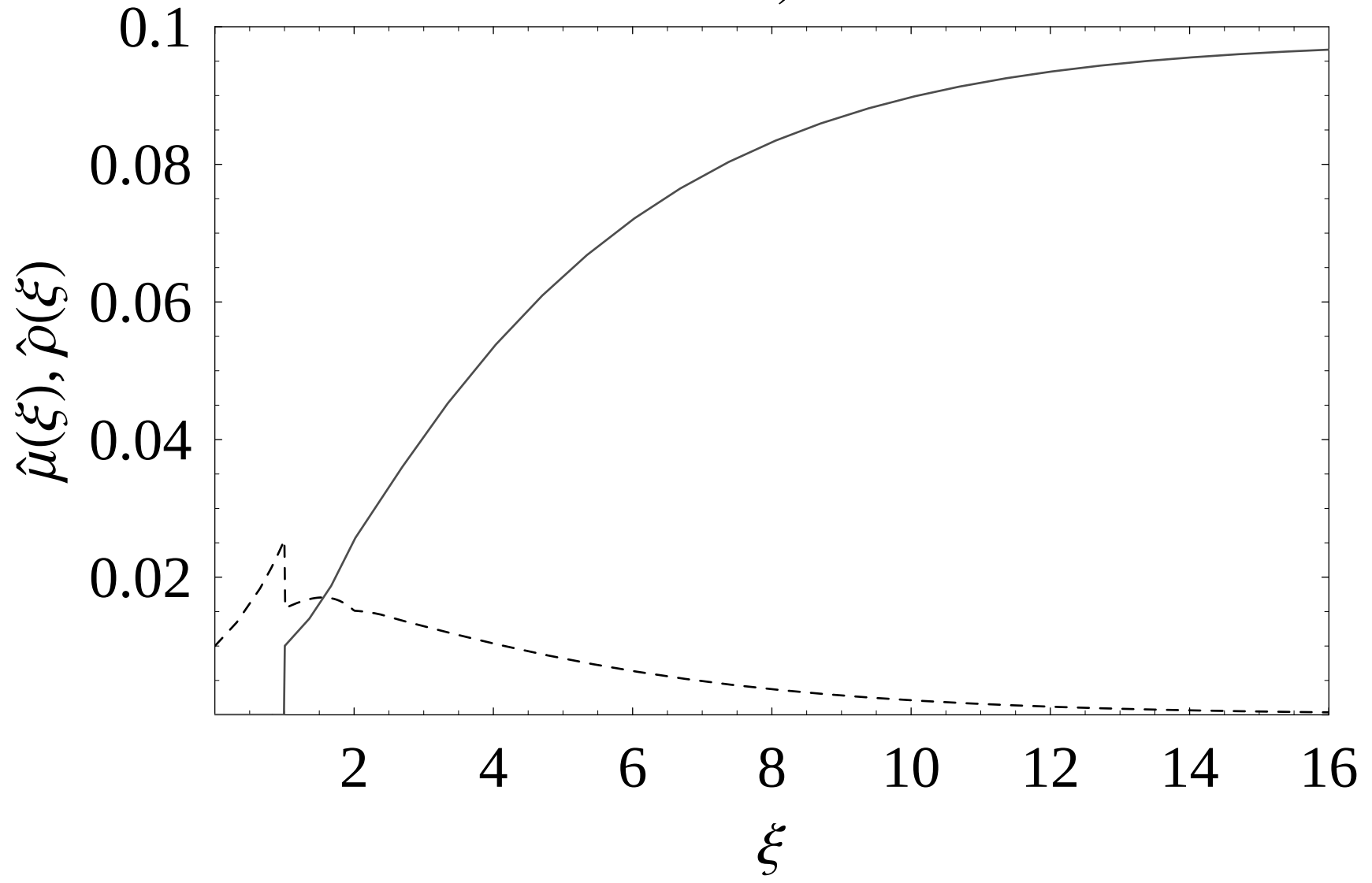

**Fig. 3**

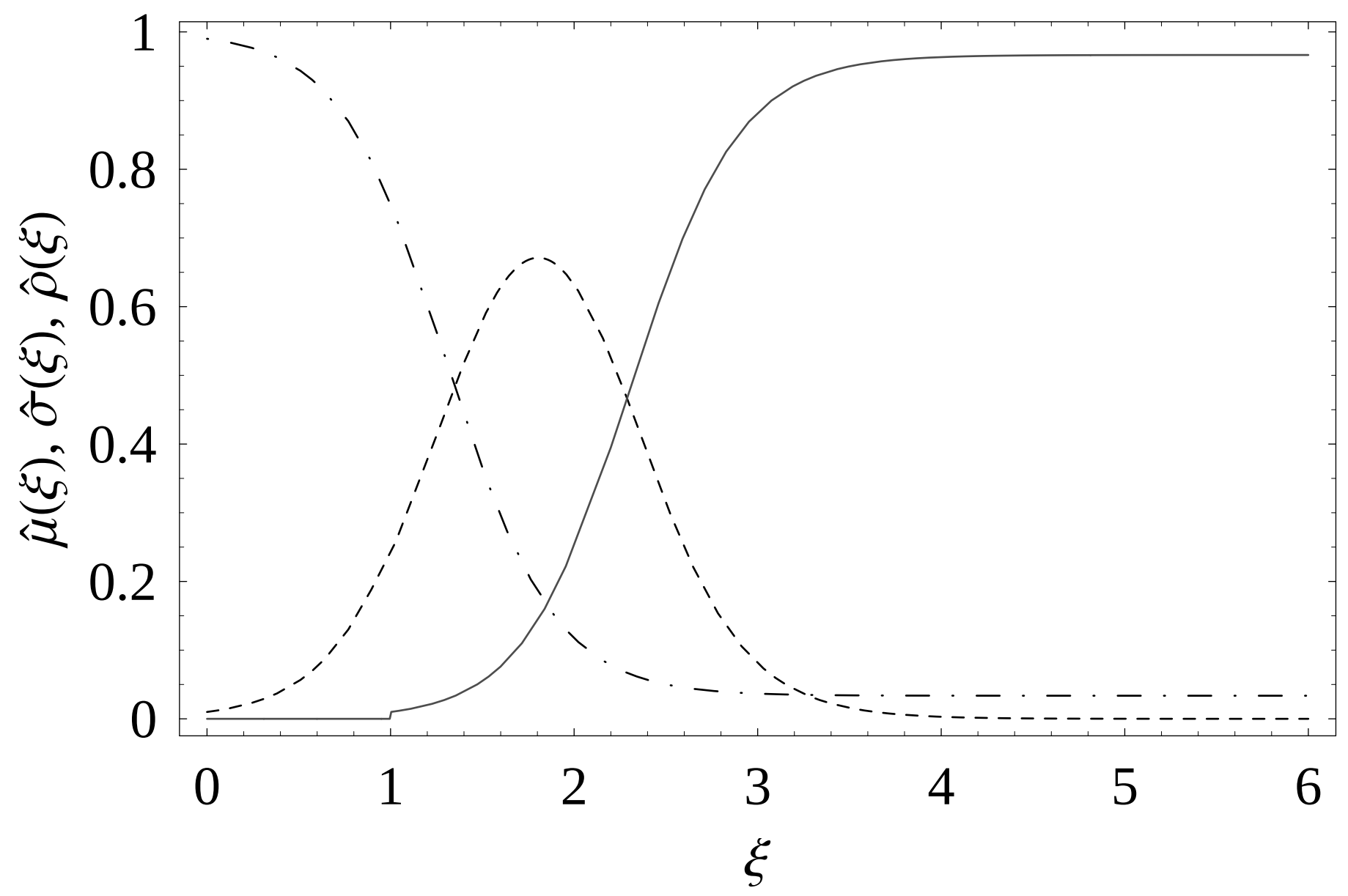

**Fig. 4a**

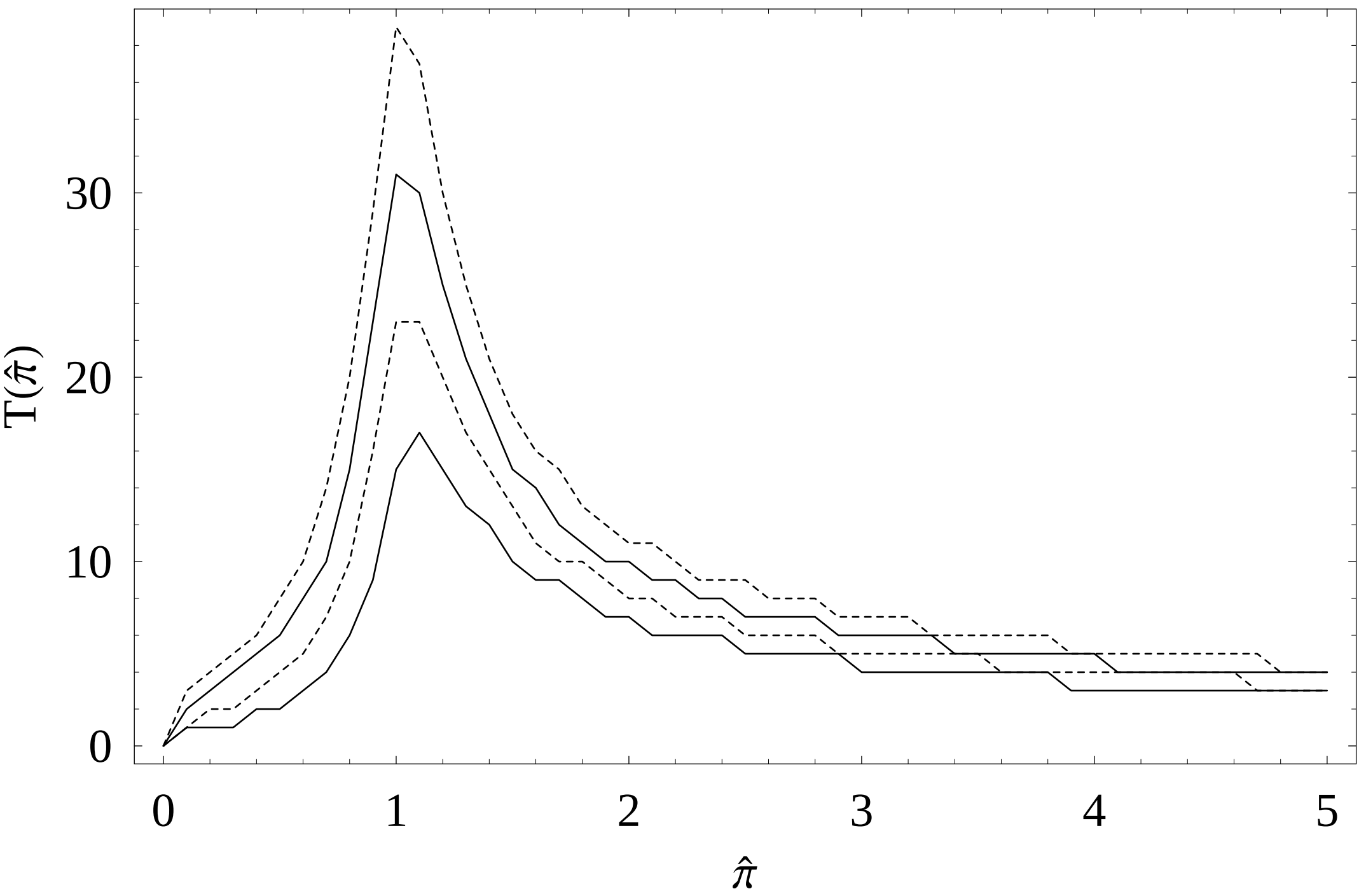

**Fig. 4b**

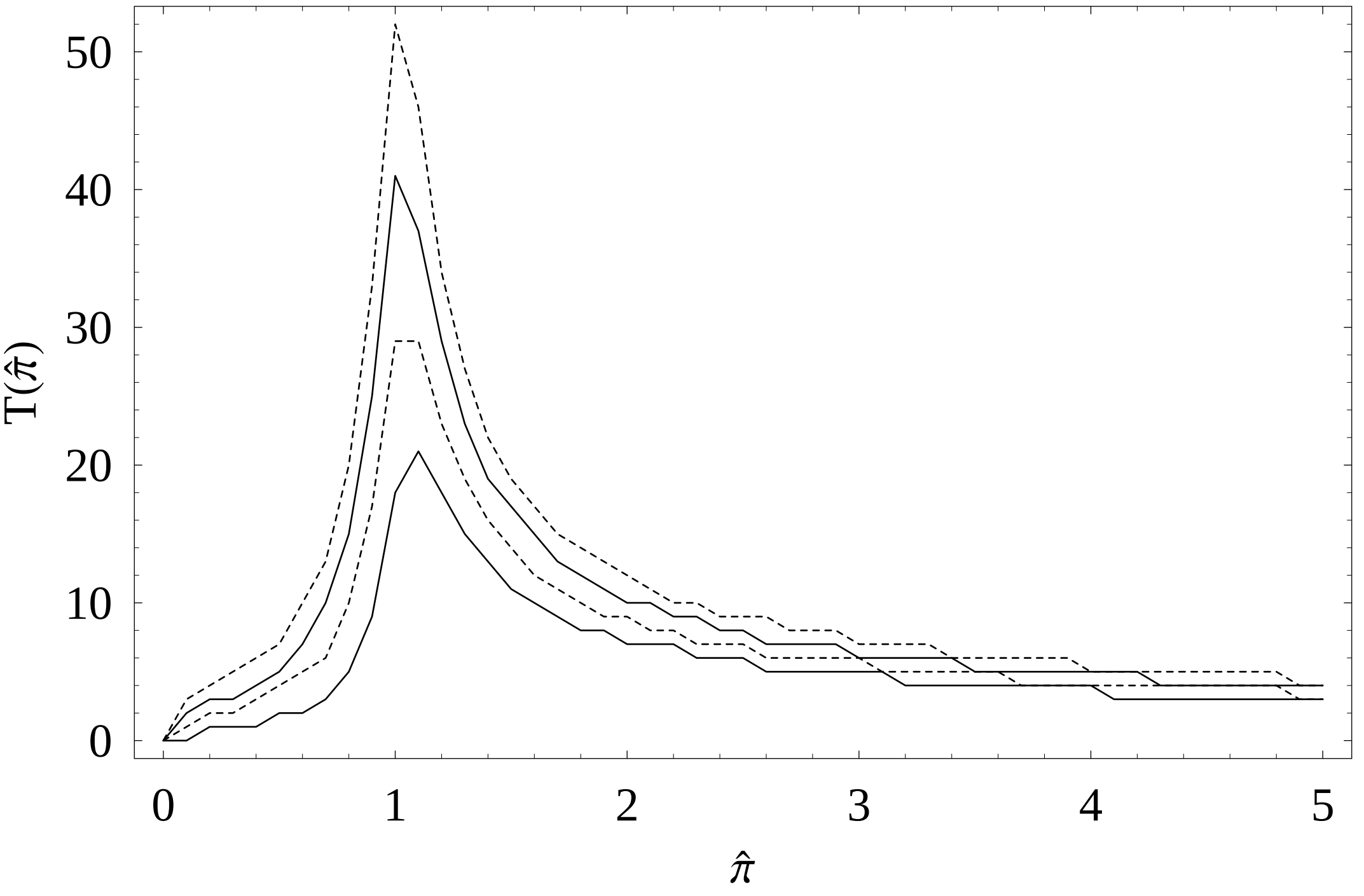

**Fig. 5a**

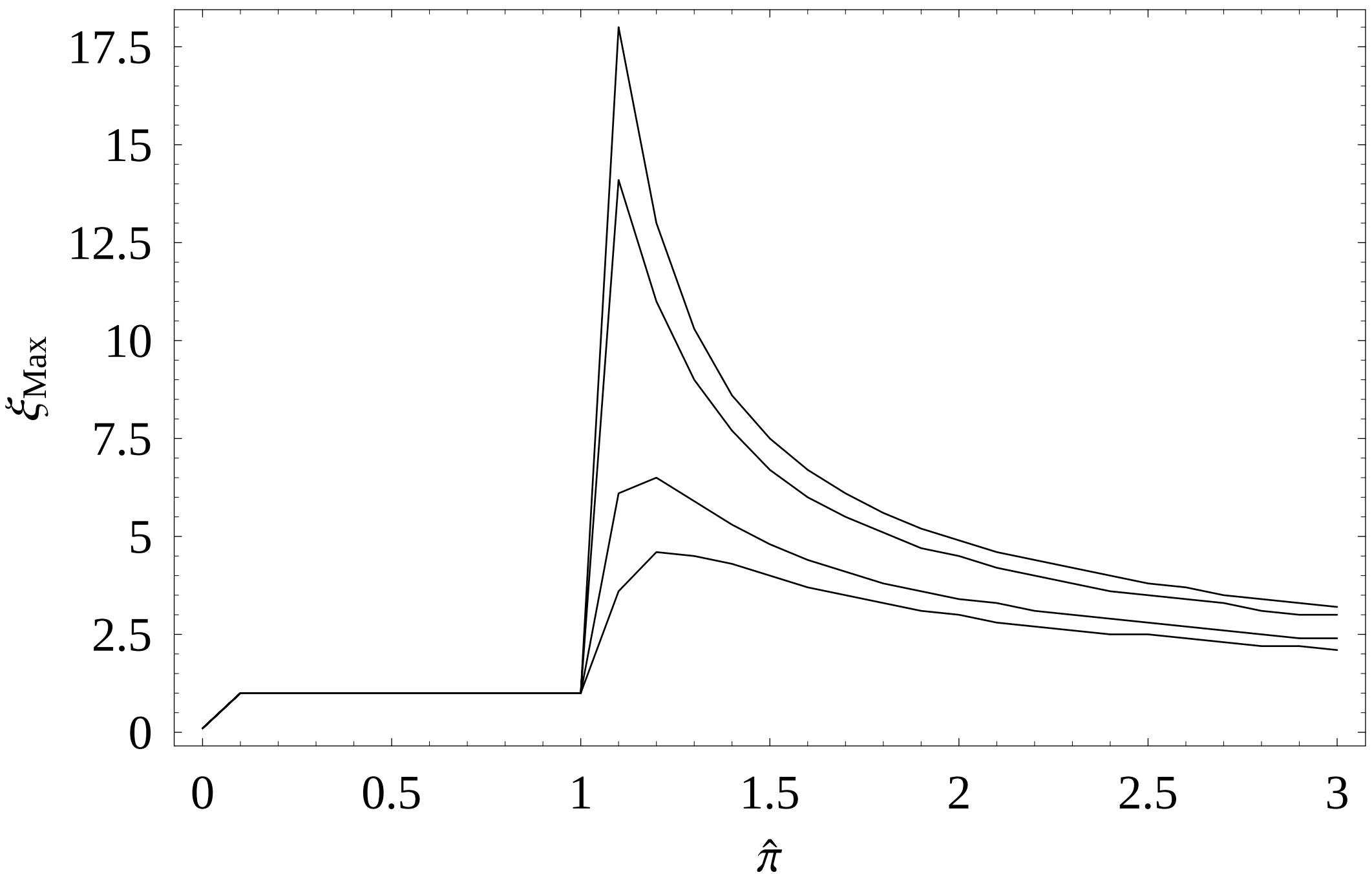

**Fig. 5b**

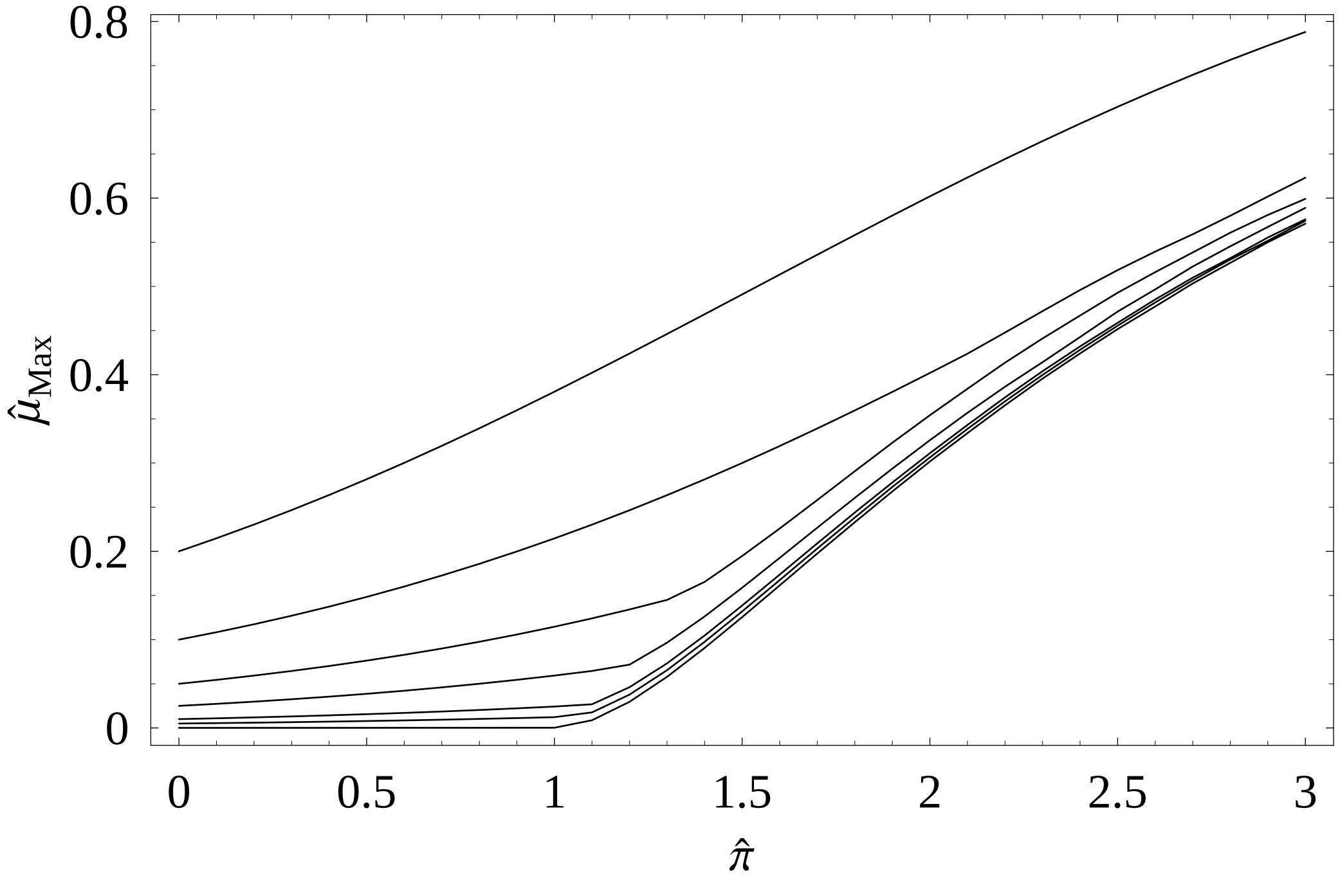

**Fig. 6**

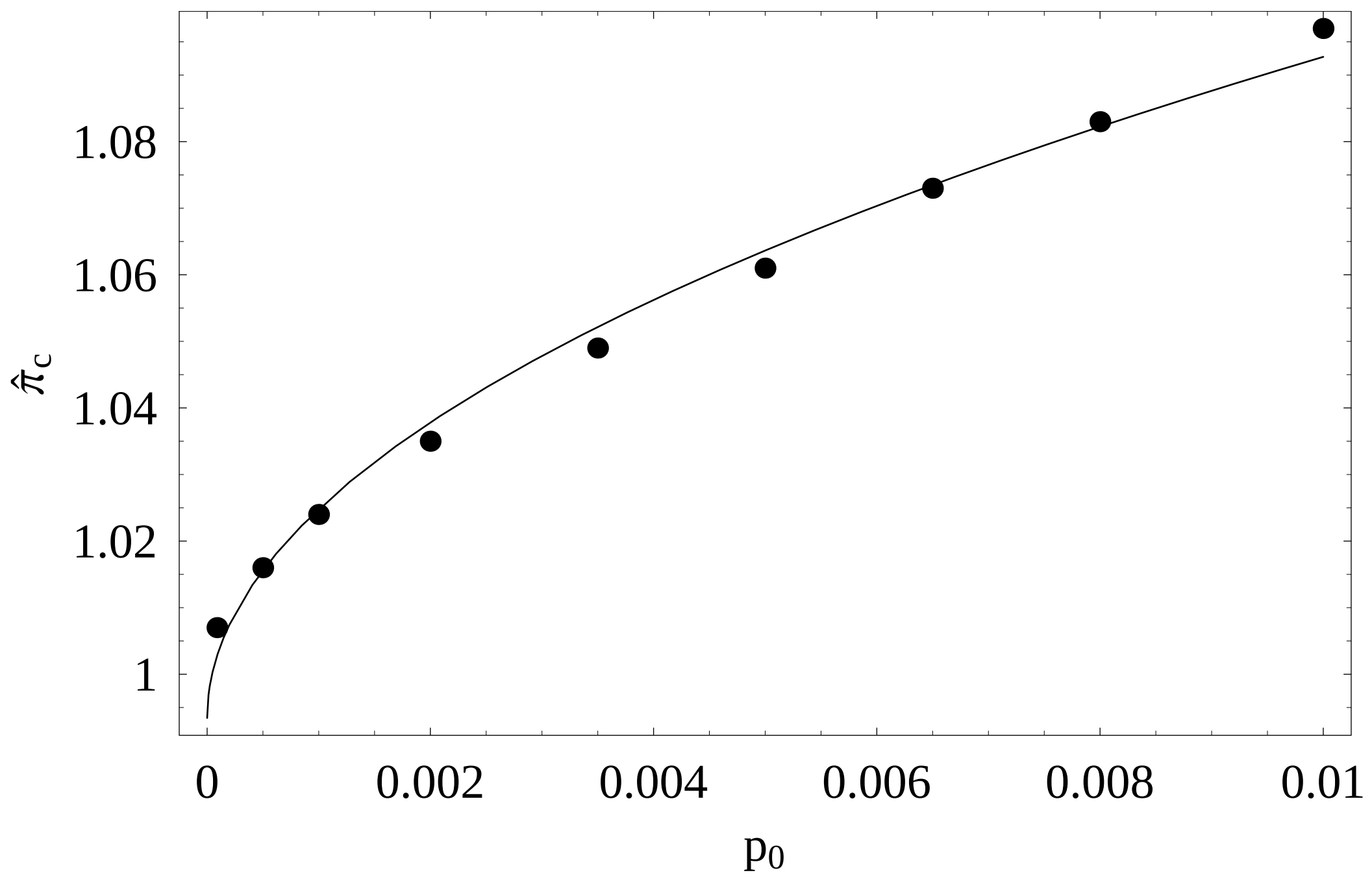